\documentclass[twoside]{article}

\usepackage{graphicx}
\usepackage{amsmath,amssymb}

\usepackage{cmpj2e}
\usepackage{cite}

\def\Gap{\Delta} 

\date{July 6, 2009}

\title[Magnetocaloric effect in quantum spin-$s$ chains]%
{Magnetocaloric effect in quantum spin-$s$ chains}
\author{A.\ Honecker\refaddr{goe}, S.\ Wessel\refaddr{stuttg}}
\addresses{
\addr{goe} Institut f\"ur Theoretische Physik,
          Georg-August-Universit\"at G\"ottingen,
          37077 G\"ottingen, Germany
\addr{stuttg} Universit\"at Stuttgart,
Institut f\"ur Theoretische Physik III,
Pfaffenwaldring 57, 70550 Stuttgart, Germany
}
\begin{document}

\maketitle

\begin{abstract}
We compute the entropy of antiferromagnetic
quantum spin-$s$ chains in an
external magnetic field using exact diagonalization and
Quantum Monte Carlo simulations. The magnetocaloric effect,
{\it i.e.}, temperature variations during adiabatic field
changes, can be derived from the isentropes. First, we focus
on the example of the spin-$s=1$ chain and show that one can
cool by closing the Haldane gap with a magnetic field.
We then move to quantum spin-$s$ chains and demonstrate
linear scaling with $s$ close to the saturation field.

In passing, we propose a new method to compute many low-lying
excited states using the Lanczos recursion.
\keywords
Quantum spin chains, magnetocaloric effect, entropy,
exact diagonalization, Quantum Monte Carlo
\pacs 
75.10.Pq; 
75.30.Sg; 
75.50.Ee; 
02.70.-c  
\end{abstract}

\section{Introduction}

The magnetocaloric effect, {\it i.e.}, a temperature change induced
by an adiabatic change of an external magnetic field was discovered
in iron
by Warburg in 1881 \cite{Warburg}. This effect also has a long history
in cooling applications. For example,
adiabatic demagnetization of paramagnetic salts was the first
method to reach temperatures below 1K \cite{GiMD33}.
Among the many experimental but also theoretical
investigations to follow, we would like to mention in particular the
computation of the field variation of the entropy in a quantum
spin-1/2 $XXZ$ chain \cite{BoFi62}. This investigation is remarkable in
so far as it is one of the very first numerical computations for
an interacting quantum many-body system. The relation of the entropy
of quantum spin chains in a magnetic field to magnetic cooling
was also noted \cite{BoNa72} and even one exact computation was
performed, albeit for a chain with a strong easy-axis anisotropy \cite{BoJo77}.

The magnetocaloric effect in quantum spin systems has recently
attracted renewed attention. The reason for this is two-fold. On the
one hand, field-induced quantum phase transitions lead to universal
responses when the applied field is varied adiabatically
\cite{ZGRS03,GR05}\footnote{The presence of precursors should, however,
be noted, see, e.g., \cite{TaYa70,BoNa72}.}.
On the other hand, it was observed that the magnetocaloric effect
is enhanced by geometric frustration
\cite{Zhito,ZhiHo,ZhiTsu04,DeRi04,SPSGBPBZ,WeHo,DeRi06,CSJ06,Gencer06,SST06,STS07,SSR07,RTCGTS07,PML09,ZH09},
promising improved efficiency in low-temperature cooling applications.
In this context, the isotropic spin-1/2 Heisenberg chain, the
isotropic and anisotropic $XY$ chain in a transverse field \cite{ZhiHo},
and the quantum Ising chain in a transverse field \cite{GR05}
were also revisited. One further example of a non-frustrated quantum
spin chain where the magnetocaloric effect was computed is the
ferrimagnetic spin-$1$-$5/2$ chain \cite{BBO07}.
Finally, the adiabatic cooling rate has recently been measured
on the spin-1/2 Heisenberg chain compound
[Cu($\mu$-C$_2$O$_4$)(4-aminopyridine)$_2$(H$_2$O)]$_n$ \cite{Tsui}
and relatedly the magnetocaloric effect of the spin-1/2 isotropic 
Heisenberg and $XXZ$ chains was computed exactly \cite{KHOT}.

In this paper we will investigate magnetocaloric properties of quantum
spin-$s$ Heisenberg chains given by the Hamiltonians
\begin{equation}
H = J\,\sum_{i=1}^L \vec{S}_i \cdot \vec{S}_{i+1} - h\,\sum_{i=1}^L S_i^z \, .
\label{eq:Ham}
\end{equation}
$J > 0$ is the antiferromagnetic exchange constant, $h$ an external
magnetic field, $L$ the length of the chain, and $\vec{S}_i$ are quantum
spin-$s$ operators at site $i$. We will use periodic boundary conditions,
{\it i.e.}, $\vec{S}_{L+1} = \vec{S}_{1}$.

In section~\ref{sec:Methods} we will first summarize the methods which we
are going to use, namely exact diagonalization (ED) and Quantum Monte
Carlo (QMC) simulations using the Stochastic Series Expansion (SSE)
framework. Then we will apply these methods to the $s=1$
chain in section~\ref{sec:S1chain}. The spin $s=1$ Heisenberg chain
is famous for the presence of the so-called `Haldane' gap at
$h=0$ \cite{Haldane1,Haldane2,WH93,GJL94}. In particular, we will
illustrate with the $s=1$ chain that one can exploit the presence of a
spin gap for cooling by adiabatic magnetization at low magnetic fields.
Next, we discuss the scaling with spin quantum number $s$ in
section \ref{sec:Sscal}. In particular, we will show that there
is a quantum scaling regime at low temperatures close to the saturation
field where one has {\em linear} scaling with $s$. Finally, in
section~\ref{sec:Summary} we will conclude with a summary.

\section{Methods}

\label{sec:Methods}

\subsection{Exact diagonalization}

\label{sec:ED}

Thermodynamic quantities can be computed using spectral representations.
For example, the entropy can be written as \cite{ZhiHo}
\begin{equation}
S = \frac{1}{T\, Z} \sum_n E_n \, {\rm e}^{-E_n/T} + \ln Z \, ,
\label{eq:SpecS}
\end{equation}
where $Z = \sum_n {\rm e}^{-E_n/T}$ is the partition function and
$E_n$ are the eigenvalues of the Hamiltonian.

In order to evaluate spectral representations like (\ref{eq:SpecS}),
one needs to diagonalize the Hamiltonian. If this is done numerically exactly,
this is called Exact Diagonalization (ED). First, one should perform
a symmetry reduction. We have used translations and $S^z$-conservation
as well as reflections and spin inversion where appropriate. One can
then use a library routine to perform a full diagonalization.
Such an approach is very much in the spirit of classic work
\cite{BoFi62,BoFi64,Bonner68}. The main technical differences of our
computations and \cite{BoFi62,BoFi64,Bonner68} are:
(i) we have exploited $SU(2)$-symmetry to reconstruct the $S^z=1$ sector
 from the $S^z \ge 2$ sectors and the spin-inversion resolved $S^z=0$ sector,
(ii) we have used improved implementations of the diagonalization routines
 \cite{GrDo89,parco07},
(iii) and we have substantially more powerful computers at our disposal.
The combination of these factors
enables us to compute full spectra for bigger systems than in the 1960s.

If we are interested only in low-energy properties, we can use
iterative diagonalization
algorithms like the Lanczos method \cite{Lanczos1950,CuWi,ZDDRV}.

The basic Lanczos algorithm \cite{Lanczos1950,CuWi,ZDDRV} for a Hermitian
matrix $H$ proceeds as follows. For a given normalized start
vector $\vec{v}_1$, one defines a sequence
of normalized Lanczos vectors $\vec{v}_j$ via the recurrence relations
\begin{eqnarray}
\beta_{j+1} \, \vec{v}_{j+1} &=& H\,\vec{v}_j - \alpha_j \,\vec{v}_j
  -\beta_{j}\,\vec{v}_{j-1} \, , \nonumber \\
\alpha_j &=& \vec{v}_j \cdot H\,\vec{v}_j \, ,
\label{eq:lancRec} \\
\beta_{j+1} &=& \vec{v}_{j+1} \cdot H\,\vec{v}_j \, , \nonumber
\end{eqnarray}
with the initial conditions $\beta_1 =0$ and $\vec{v}_0 = \vec{0}$.
The real coefficients $\alpha_i$ and $\beta_i$ define the
so-called Lanczos matrices
\begin{equation}
T_n = \begin{pmatrix}
\alpha_1 & \beta_2 & 0 & \ldots & \ldots & 0 \\
\beta_2 & \alpha_2 & \ddots & \ddots & & \vdots \\
0 & \ddots & \ddots & \ddots & \ddots & \vdots \\
\vdots & \ddots & \ddots & \ddots & \ddots &  0 \\
\vdots &  & \ddots & \ddots & \ddots & \beta_n \\
0 & \ldots & \ldots & 0 & \beta_n & \alpha_n
\end{pmatrix}  \, .
\label{eq:LancMat}
\end{equation}
The crucial point of the Lanczos algorithm is that the
eigenvalues of $T_n$ yield a good approximation
of the extremal eigenvalues of $H$ already for
$n$ much smaller than the total Hilbert space dimension.

The Lanczos algorithm suffers from one practical
problem: the recursion relations (\ref{eq:lancRec})
are supposed to guarantee mutual orthogonality of
the vectors $\vec{v}_i$, {\it i.e.},
$\vec{v}_i \cdot \vec{v}_j = 0$ for $i \ne j$.
However, this fails to be correct when the computations
are carried out numerically, leading to undesired
spurious states, called ghost states \cite{CuWi,ZDDRV}.
Several strategies have been proposed \cite{CuWi,ZDDRV} in
order to deal with the ghost problem during the computation
of excited eigenvalues.

Here we suggest another method to compute excited states with
controlled accuracy and correct multiplicity. We start by performing
a fixed number $n$ of Lanczos iterations with a given start
vector $\vec{v}_1$. Next we compute the orthogonal transformation
$U_{i,j}$, $1 \le i,j \le n$ diagonalizing the Lanczos matrix
(\ref{eq:LancMat}). Then we fix a number $m \ll n$ and repeat
the Lanczos procedure with the same start vector $\vec{v}_1$.
During this second Lanczos pass we construct $m$ vectors
\begin{equation}
\vec{u}_i = \sum_{j=1}^n U_{j,i} \, \vec{v}_j \, , \qquad 1 \le i \le m \, .
\label{eq:LancApproxEv}
\end{equation}
The second Lanczos pass is needed in order to avoid storing the
$n \gg m$ Lanczos vectors $\vec{v}_j$.

The vectors $\vec{u}_i$ given by (\ref{eq:LancApproxEv}) yield
approximations to the eigenvectors of $H$. However, due to the
ghost problem, there are spurious vectors which need to be eliminated.
We perform this in two steps. First, using the observation that
the vectors $\vec{u}_i$ are supposed to be mutually orthogonal,
we can eliminate those $\vec{u}_i$ which have big projections
on the $\vec{u}_j$ with $j < i$. Second, we reorthogonalize
the remaining vectors yielding $\tilde{m} \le m$ orthonormal
vectors $\tilde{\vec{u}}_i$. Finally, we project the matrix $H$
onto this subspace via
\begin{equation}
H_{i,j} = \tilde{\vec{u}}_i \cdot H \, \tilde{\vec{u}}_j \, .
\label{eq:LancProjH}
\end{equation}
A full diagonalization of the $\tilde{m} \times \tilde{m}$ matrix
$H_{i,j}$ and application of the resulting basis transformation
to the vectors $\tilde{\vec{u}}_i$ yields $\tilde{m}$ {\em orthogonal}
vectors $\vec{w}_i$ which approximate the eigenvectors of $H$.
The  eigenvalues $E_i$  associated to the
vectors $\vec{w}_i$ and their accuracy can be estimated according to
\begin{equation}
E_i = \vec{w}_i \cdot H \, \vec{w}_i \, , \qquad
\delta E_i^2 = \vec{w}_i \cdot \left( H - E_i \right)^2 \vec{w}_i \, .
\label{eq:DeltaEi}
\end{equation}

According to our experience, there are neither any missing nor any spurious
eigenvalues among the converged ones. It should be mentioned that there is
no guarantee that the Lanczos procedure yields the complete spectrum for
a given start vector $\vec{v}_1$, in particular if one has not performed
a complete symmetry decomposition. Nevertheless, numerical noise seems to
prevent this from happening, at least for sufficiently generic start
vectors $\vec{v}_1$.

For our procedure one needs to choose the
number of Lanczos iterations $n$ and the dimension of the subspace $m \le n$
in advance and the whole procedure needs to be repeated if it turns
out that less than the desired number of converged eigenvalues have
been obtained. In practice, however, one obtains results of a comparable
quality for a class of problems where the choice of $n$ and $m$ has
been adjusted for one representative case. In the examples to be reported
below, we have been able to obtain about 100 eigenvalues in a given
symmetry sector with a relative accuracy of $10^{-8}$ or better
using $n=2000$ Lanczos iterations and $m=500$ vectors $\vec{u}_i$.

\subsection{Quantum Monte Carlo}

\label{sec:QMC}

Since the system sizes accessible by either full diagonalization
or Lanczos diagonalization are limited, it is desirable to have
other methods at our disposal. For non-frustrated spin models
like the spin-$s$ Heisenberg chain (\ref{eq:Ham}), one can in
principle use Quantum Monte Carlo (QMC) simulations. There is,
however, one problem: we are particularly interested in
the entropy which is usually obtained with large statistical errors
from integrating a Monte-Carlo result for the specific heat.
In this subsection we will summarize how one can
circumvent this problem.

In order to obtain a QMC estimate of the temperature dependence of the entropy,
we employed an extended ensemble, broad-histogram method~\cite{QWL,QuantumHistogramMethods,opti}
within the Stochastic Series Expansion (SSE) framework~\cite{SSE,OperatorLoops,DirectedLoops,GenDirLoop}.
Based on this approach, thermodynamic quantities can be obtained over a broad range of temperatures from a single 
QMC simulation
that provides estimates of the expansion coefficients $g(n)$ of the system's
partition function $Z$ in a high-temperature series expansion:
\begin{equation}
Z=\mathrm{Tr} \, {\rm e}^{-\beta H}=\sum_{n=0}^\infty g(n) \beta^n.
\end{equation}
Here, $\beta=1/T$ is the inverse temperature. In terms of the expansion coefficients $g(n)$,
the internal energy is obtained using
\begin{equation}
E=\langle H \rangle = - \frac{\partial}{\partial \beta} \ln Z
= - \frac{1}{Z} \sum_{n} g(n) n \beta^{n-1} = - \frac{1}{\beta} \langle n \rangle ,
\end{equation}
and the free energy from
\begin{equation}
F=-\frac{1}{\beta}\ln{Z}=-\frac{1}{\beta}\ln \sum_n g(n)\beta^n.
\end{equation}
Finally, the entropy can be calculated using $S=(E-F)/T$.

In the QMC simulation, estimates of the first $\Lambda$ coefficients $g(n), n=0,...,\Lambda$ are
obtained by performing a random walk in the expansion order $n$, such as to sample efficiently all expansion orders 
from
$n=0$ (where $g(0)=(2s+1)^L$ for a spin-$s$ chain of $L$ sites is known exactly)
to $n=\Lambda$. This is accomplished using an extension of the Wang-Landau flat-histogram sampling
algorithm~\cite{WangLandau1,WangLandau2} to the quantum case, as detailed previously
in Refs.~\cite{QWL,QuantumHistogramMethods,opti}.
Knowledge of the first $\Lambda$ coefficients $g(n)$ allows for the calculation of thermodynamic quantities
from $\beta=0$ down to a
temperature $T_m=1/\beta_m$ for a system of $L$ sites, where $\Lambda$ scales proportional to $\beta_m\times 
L$~\cite{QWL}. 
In particular, for the 
spin $s=1$ Heisenberg chain, we were able to treat systems with up to $L=40$ sites,
accessing temperatures down to $T_m \approx 0.03$, which required an already large value of $\Lambda=5000$.
The data shown in this contribution (and also those in \cite{WeHo})
were obtained based on the original version of the algorithm~\cite{QWL},
while recently an improved sampling strategy was proposed,
based on optimizing the broad histogram Quantum Monte Carlo ensemble~\cite{opti}.

\subsection{A test case}

\label{sec:test}

\begin{figure}[t!]
\begin{center}
\includegraphics[width=0.6\columnwidth]{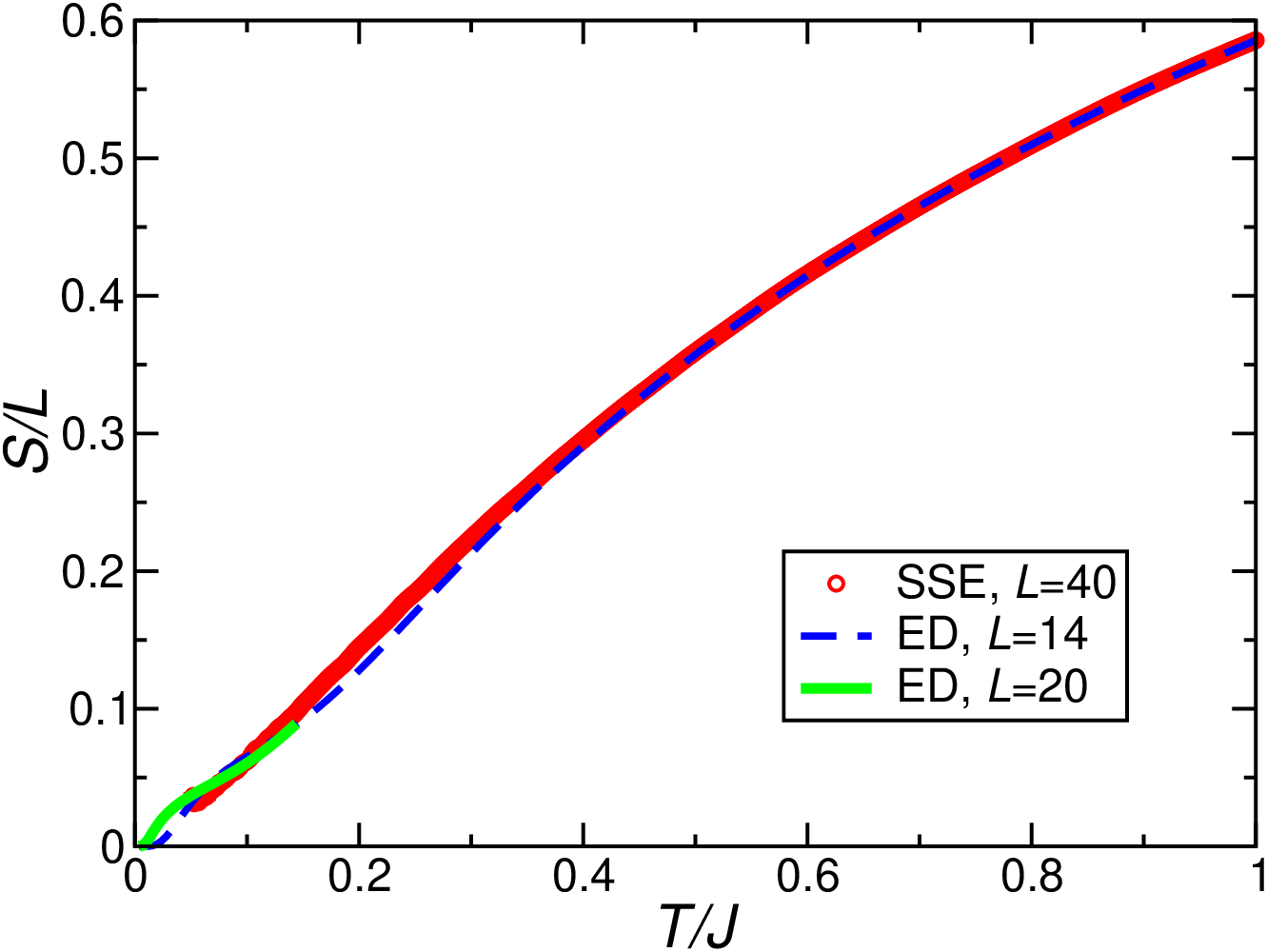}
\end{center}
\caption{Entropy per site $S/L$ of the $s=1$ Heisenberg
chain as a function of temperature $T$ for $h=3\,J$.
Lines are obtained by ED for $L=14$ and $20$ sites,
symbols show QMC results for $L=40$ sites.
}
\label{figS1entH3}
\end{figure}

We test and compare the aforementioned numerical methods using the
example of the $s=1$ Heisenberg chain (\ref{eq:Ham}) with
a finite magnetization.
Fig.~\ref{figS1entH3} shows the entropy per site $S/L$
as a function of temperature $T$
for a fixed magnetic field $h=3\,J$. For $L=14$, we have performed
a full diagonalization. The corresponding ED curve can therefore
be considered as the exact result for $L=14$.

A full determination of the spectrum is clearly out of reach
already for $L=20$. In this case, we therefore had to perform
a severe truncation to low energies. Thus, the $L=20$ ED curve
is only a low-temperature approximation. In this paper, the
precise temperature range is fixed as follows: Let ${\cal E}$ be the
highest energy until which the spectrum is definitely complete.
Then we restrict to temperatures $T\le {\cal E}/5$. This choice
ensures that missing
states are suppressed by a Boltzmann factor $\le \exp(-5) \approx
0.006738$. This may seem a small number, but since we are discarding
many states, extending truncated data to higher temperatures would
yield artifacts which are clearly visible on the figure(s).

QMC (denoted by SSE in Fig.~\ref{figS1entH3}) is able to treat the larger
system size $L=40$. This method, however, is best suited
for high temperatures. Indeed, at $T \approx 0.05\,J$ one
can see deviations caused by statistical errors in the
SSE result of Fig.~\ref{figS1entH3} which prevented us
from going to lower temperatures. Finite-size effects
can be observed in Fig.~\ref{figS1entH3} in the
$L=14$ ED curve for $T \lesssim 0.4\,J$ and in the
$L=20$ ED result. Otherwise
it is reassuring that we observe good overall agreement between
all three methods.

In passing we note that, for the value of the magnetic field
used in Fig.~\ref{figS1entH3}, the low-energy physics of
the $s=1$ chain is described by a Luttinger liquid (see
Ref.~\cite{Fath03} and references therein). It is
well known that the specific heat $C$ of a Luttinger
liquid is linear in $T$ \cite{Giamarchi}. Due to the
relation $C=T\,(\partial\,S/\partial T)$ and because
of $S(T=0) = 0$, the entropy
of a Luttinger liquid is identical to its specific heat
and in particular also linear in $T$. Indeed,  Fig.~\ref{figS1entH3}
is consistent with a linear
behavior $S \propto T$ at sufficiently large $L$ and low $T$.

\section{Spin $s=1$ Heisenberg chain}

\label{sec:S1chain}

\begin{figure}[t!]
\begin{center}
\includegraphics[width=0.6\columnwidth]{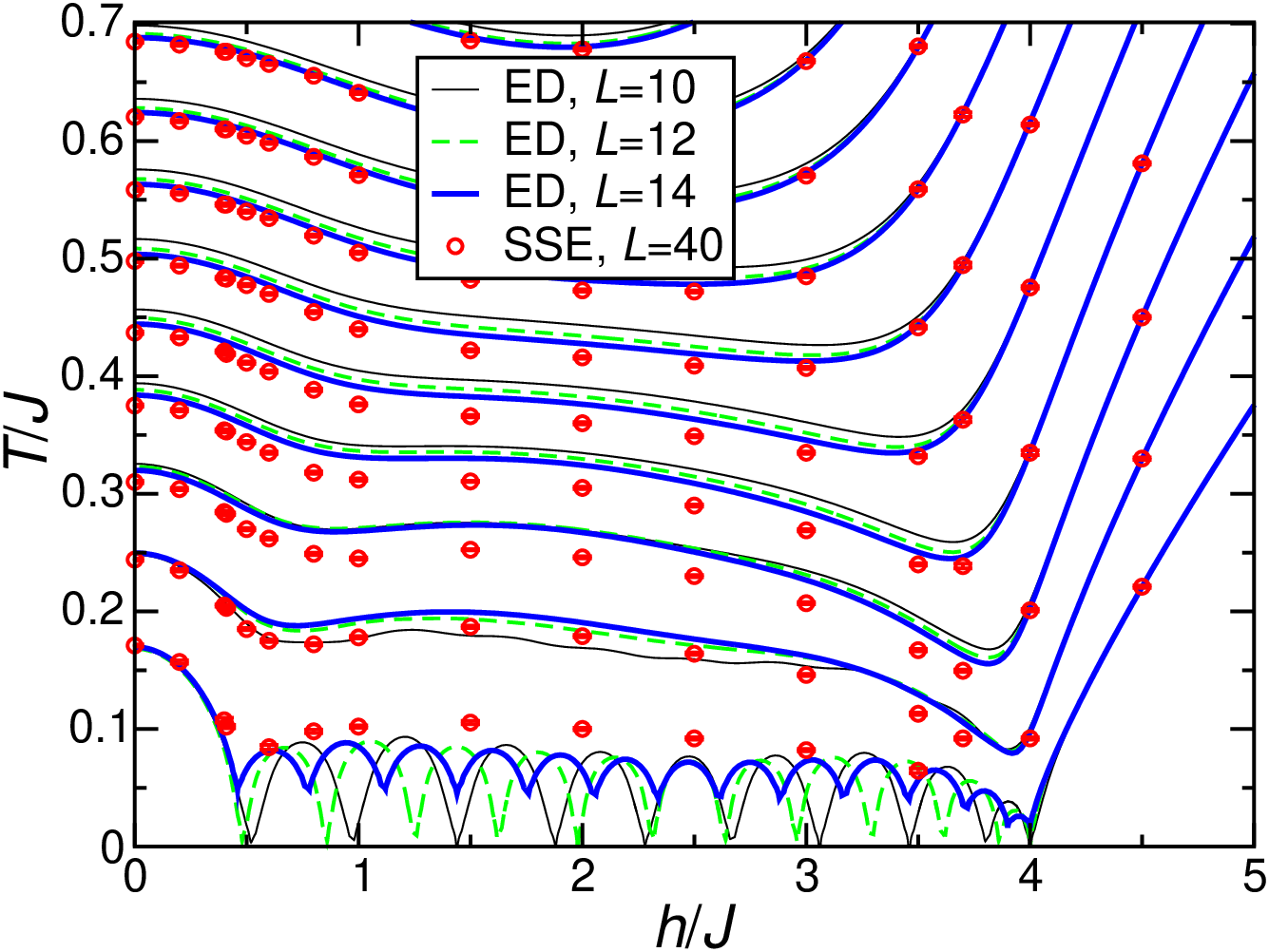}
\end{center}
\caption{Isentropes, {\it i.e.}, adiabatic demagnetization curves
of the $s=1$ Heisenberg chain as obtained
from ED and QMC for different chain lengths.
The corresponding values of the entropy are
$S/L = 0.05$, $0.1$, $0.15$, $0.2$, $0.25$, $\ldots$ (bottom to top).
}
\label{figS1isent}
\end{figure}

In this section we will discuss the entropy of the $s=1$ Heisenberg chain
in more detail.

First, we show results for the value of the entropy per
site $S/L$ as a function of $h$ and $T$
in Fig.~\ref{figS1isent}. The ED curves have been
obtained by computing the entropy $S$ on a mesh in the $h$-$T$-plane
and determining the constant entropy curves, {\it i.e.},
the isentropes from this data. This leads to some discretization
artifacts at the cusps in the the $S/L=0.05$ curves (the ED curves
at the lowest temperature) in Fig.~\ref{figS1isent}. Conversely,
the QMC had problems to resolve the large low-temperature entropy
around the saturation field $h_{\rm sat.} = 4\,J$. Accordingly,
the $L=40$ SSE data points with $S/L=0.05$ are missing at
$h/J=3.7$ and $4$. From our
QMC simulations we can just conclude that for an $L=40$ chain
these two points 
are in the region $T < 0.067\,J$ and $T < 0.085\,J$ for
$h=3.7\,J$ and $4\,J$, respectively.

The ED curves with $S/L=0.05$ in Fig.~\ref{figS1isent} have
marked finite-size wiggles. Also at higher temperatures
(entropies), finite-size effects can be observed. There are, however,
two regions where finite-size effects are evidently small. One
is the low-temperature regime for fields $h$ smaller than the
Haldane gap \cite{WH93,GJL94}
\begin{equation}
\Gap \approx 0.4105\,J \, .
\label{eq:HaldaneGap}
\end{equation}
Also in the gapped high-field regime above the saturation field
$h \gtrsim 4\,J$ one observes essentially no finite-size effects
at any temperature. This renders ED the method of choice at high
fields $h \gtrsim h_{\rm sat.}$, in particular at low temperatures,
while QMC is preferable otherwise because bigger system sizes $L$
are accessible.

One can read off from Fig.~\ref{figS1isent} pronounced temperature changes 
in two regimes. Firstly, one can obtain cooling by adiabatic demagnetization
from a high magnetic field $h > h_{\rm sat.}$ as one lowers the field
to the saturation field. Secondly, at low temperatures, one can also cool
by adiabatic magnetization from $h=0$ to $h=\Gap$. This demonstrates
that one can use the gap-closing at a generic field-induced quantum phase
transition for cooling purposes. In the regime
$0.4105\,J \approx \Gap < h < h_{\rm sat.} = 4\,J$, the spectrum is
gapless (compare Fig.~\ref{figS1entH3} and the related discussion
in section \ref{sec:test}).
Accordingly, one observes only small temperature changes induced
by adiabatic (de)magnetization in this field range.

\begin{figure}[t!]
\begin{center}
\includegraphics[width=0.6\columnwidth]{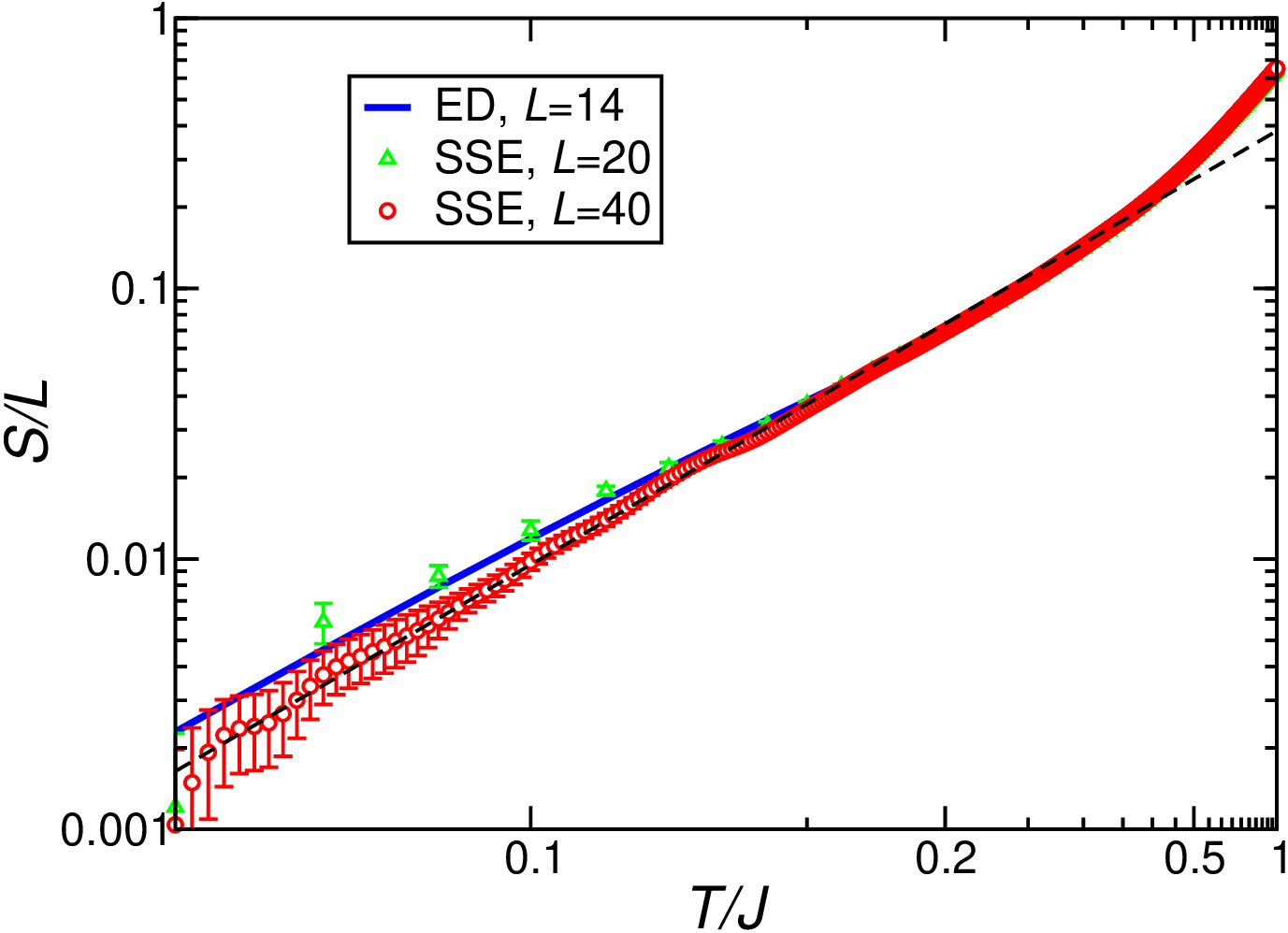}
\end{center}
\caption{Entropy per spin of the $s=1$ Heisenberg chain at zero magnetic
field $h=0$ as a function of temperature. Note that the temperature
axis is scaled as $1/T$ and the entropy axis logarithmically.
The dashed line is a fit to (\ref{eqEntGap}) with ${\cal A} = 0.577$
and $\Gap = 0.4105\,J$.}
\label{figS1entH0}
\end{figure}

We now focus on the low-field regime. Firstly, we consider the
behavior of the entropy at $h=0$ which is shown in Fig.~\ref{figS1entH0}.
One observes that finite-size effects are small for $L \ge 14$:
the $L=14$ curve is almost within the error
margins of the $L=40$ QMC curve even at low temperatures.
This can be attributed to a finite correlation length $\xi$
which is evidently sufficiently smaller than $14$ for all
temperatures at $h=0$. Indeed, the correlation length
of the $s=1$ Heisenberg chain is known to be
$\xi \approx 6$ at $h=0$ and $T=0$ \cite{WH93,GJL94}.

Because of the presence of a gap $\Gap$ at $h=0$, we expect the entropy
to be exponentially activated as a function of temperature. Accordingly,
the low-temperature asymptotic behavior at $h=0$ should follow the form
\begin{equation}
\left.\frac{S}{L}\right|_{h=0} = {\cal A} \, {\rm e}^{-\Gap/T} \, .
\label{eqEntGap}
\end{equation}
If we fix the gap to the known value $\Gap = 0.4105\,J$ \cite{WH93,GJL94},
the only free parameter in the low-temperature asymptotic behavior
(\ref{eqEntGap}) is the prefactor ${\cal A}$. A fit of the $L=40$
QMC results yields
${\cal A} = 0.577(1)$. The dashed line in Fig.~\ref{figS1entH0} shows
that the formula (\ref{eqEntGap}) describes the low-temperature regime
quite well with the given parameters.

\begin{figure}[t!]
\begin{center}
\includegraphics[width=0.6\columnwidth]{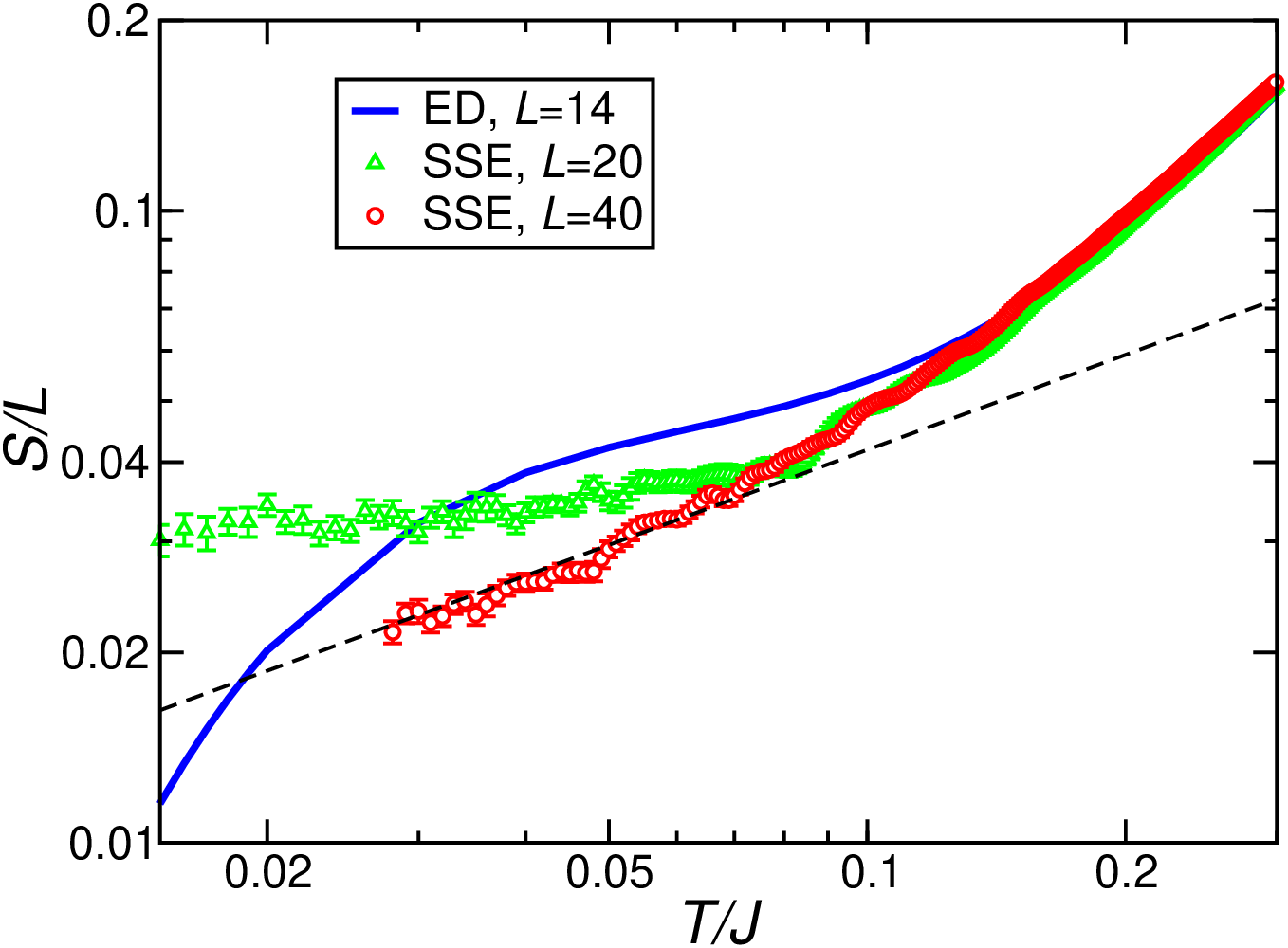}
\end{center}
\caption{Entropy per spin of the $s=1$ Heisenberg chain at
$h=0.4105\,J$, corresponding to the Haldane gap.
Note the doubly logarithmic scale.
The dashed line shows the square-root behavior (\ref{eqEntSqrt})
with ${\cal B} = 0.1323$.}
\label{figS1entHgap}
\end{figure}

Secondly, we consider a magnetic field exactly equal
to the Haldane gap $h=0.4105\,J$. Fig.~\ref{figS1entHgap}
shows a doubly-logarithmic plot of the entropy $S$ as
a function of temperature $T$ for this value of the magnetic
field. Since we are now sitting at a field-induced quantum
phase transition, we expect a large number of low-lying
states and relatedly an infinite correlation length $\xi = \infty$.
On a finite system, these low-lying states will be pushed to higher
energies. Indeed, we observe pronounced finite-size effects
at low temperatures in Fig.~\ref{figS1entHgap}.

At a quantum phase transition in one dimension which preserves
a $U(1)$-symmetry, the entropy $S$ is expected \cite{BoNa72,ZhiHo,ZGRS03,GR05}
to vary asymptotically as a square root of temperature:
\begin{equation}
\left.\frac{S}{L}\right|_{h=\Gap} = {\cal B} \, \sqrt{\frac{T}{J}} \, .
\label{eqEntSqrt}
\end{equation}
As before, the amplitude ${\cal B}$ is the only free parameter.
A fit to the $L=40$ QMC data with $T\le 0.07\,J$ yields
${\cal B}=0.1323(6)$. The dashed line in Fig.~\ref{figS1entHgap}
shows that the $L=40$ QMC data is indeed consistent with
the asymptotic square root (\ref{eqEntSqrt})
in the range $0.03 \lesssim T/J \lesssim 0.07$. However, because
of finite-size effects, in this case we really need to go to $L=40$
to recover the asymptotic behavior. Furthermore, we need to
restrict to $T\lesssim 0.07\,J$ since one can see from Fig.~\ref{figS1entHgap}
that a pure power law is no longer a good description for
$T \gtrsim 0.07\,J$.

Comparison of (\ref{eqEntGap}) and (\ref{eqEntSqrt}) shows that one
can reach exponentially small temperatures $T_{\rm f}$ at $h=\Gap$
by adiabatic magnetization of a spin-$1$ Heisenberg chain from
an initial temperature $T_{\rm i}$ at $h=0$. For an initial temperature
$T_{\rm i} \lesssim 0.1\,J$, the final temperature $T_{\rm f}$ can be
quantitatively estimated from the following combination of
(\ref{eqEntGap}) and (\ref{eqEntSqrt})
\begin{equation}
\frac{T_{\rm f}}{J}
 = \frac{{\cal A}^2}{{\cal B}^2} \, {\rm e}^{-2\,\Gap/T_{\rm i}} \, ,
\label{eqS1Tf}
\end{equation}
with the parameters $\Delta$, ${\cal A}$, and ${\cal B}$ given above.

\section{Scaling with spin quantum number $s$}

\label{sec:Sscal}

We will now turn to the high-field region and consider in particular
scaling with the spin quantum number $s$ close to the saturation field.

\begin{figure}[t!]
\begin{center}
\includegraphics[width=0.6\columnwidth]{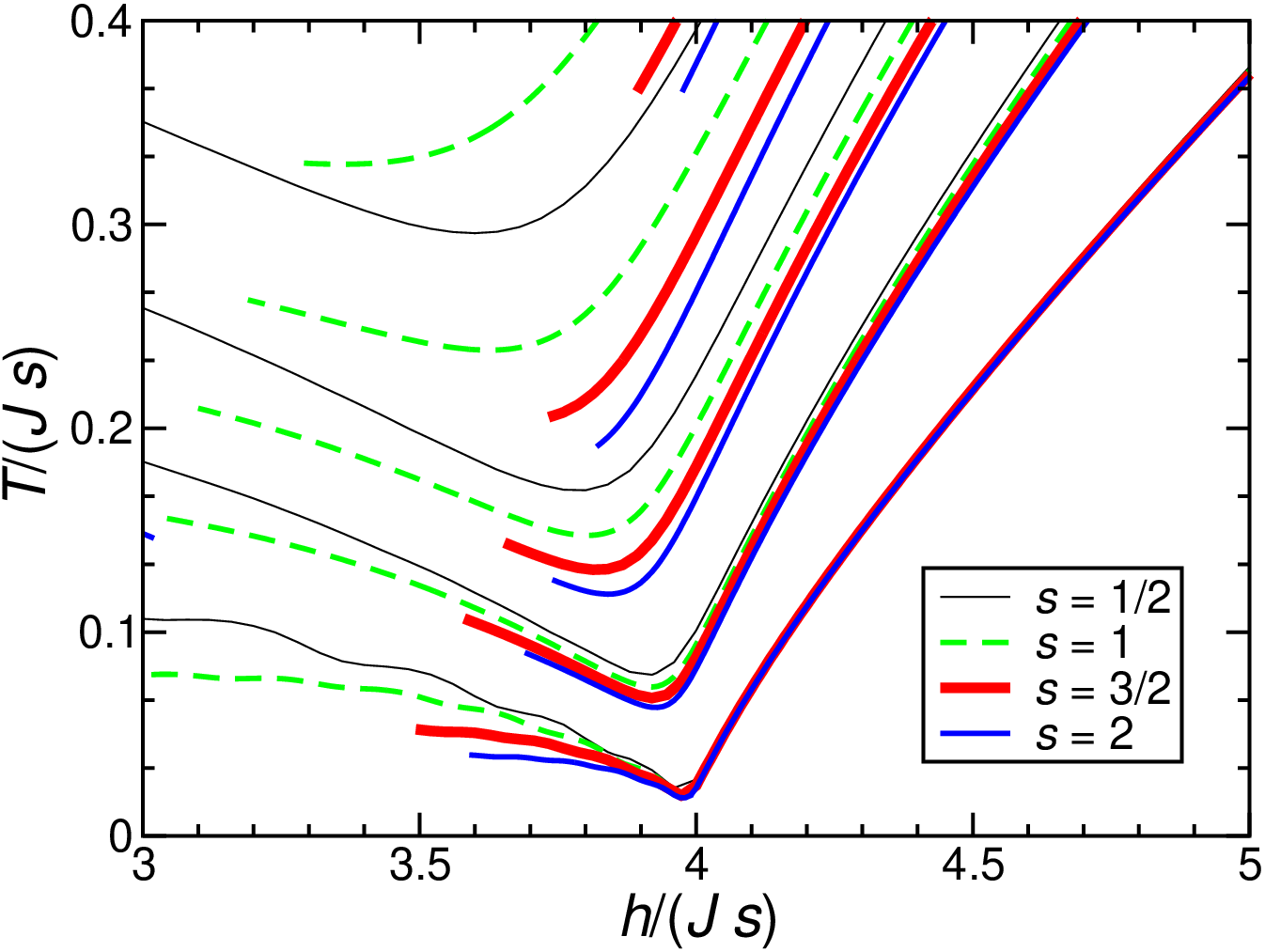}
\end{center}
\caption{Isentropes of $L=20$ Heisenberg chains as obtained
by ED for different $s$. Note that a truncation of the
energy spectrum has been used for $s > 1/2$. The
values of the entropy per site are
$S/L=0.05$, $0.01$, $0.15$, $0.2$, and $0.25$
(bottom to top).}
\label{figMad}
\end{figure}

It is straightforward to compute the excitation energy
$\varepsilon$ of a single
flipped spin propagating above a ferromagnetically polarized background
(see, e.g., \cite{HoPa85}):
\begin{equation}
\varepsilon(k) = 2\,J\,s\,\left(\cos k-1\right) + h \, .
\label{eq:singleMagnon}
\end{equation}
The minimum of this one-magnon dispersion is located at $k=\pi$.
The saturation field $h_{\rm sat.}$ of the spin-$s$ Heisenberg chain
(\ref{eq:Ham}) is given by the one-magnon instability \cite{PaBo85}.
Accordingly, it can be determined by inserting the condition
$\varepsilon(\pi) = 0$ into (\ref{eq:singleMagnon}):
\begin{equation}
h_{\rm sat.} = 4\,J\,s \, .
\label{hsat}
\end{equation}

The fact that the single-particle energy (\ref{eq:singleMagnon})
scales linearly with $s$ suggests to scale all energies 
close to the saturation field and at low temperatures linearly
with $s$. In order to test this scenario, we have first computed
the isentropes for a fixed chain length $L=20$
and $s \le 2$. The $s=1/2$
curves are obtained by a full diagonalization of the Hamiltonian \cite{ZhiHo}
while we have performed additional ED computations for $s=1$, $3/2$,
and $2$ using the truncation procedure described in section \ref{sec:Methods}.
Fig.~\ref{figMad} shows the resulting isentropes for
$S/L=0.05$, $0.01$, $0.15$, $0.2$, and $0.25$ close
to the saturation field (\ref{hsat}). We observe that
at low temperatures,
scaling both $h$ and $T$ by $s$ leads to a nice collapse
of the isentropes around and in particular above the
saturation field.

\begin{figure}[t!]
\begin{center}
\includegraphics[width=0.48\columnwidth]{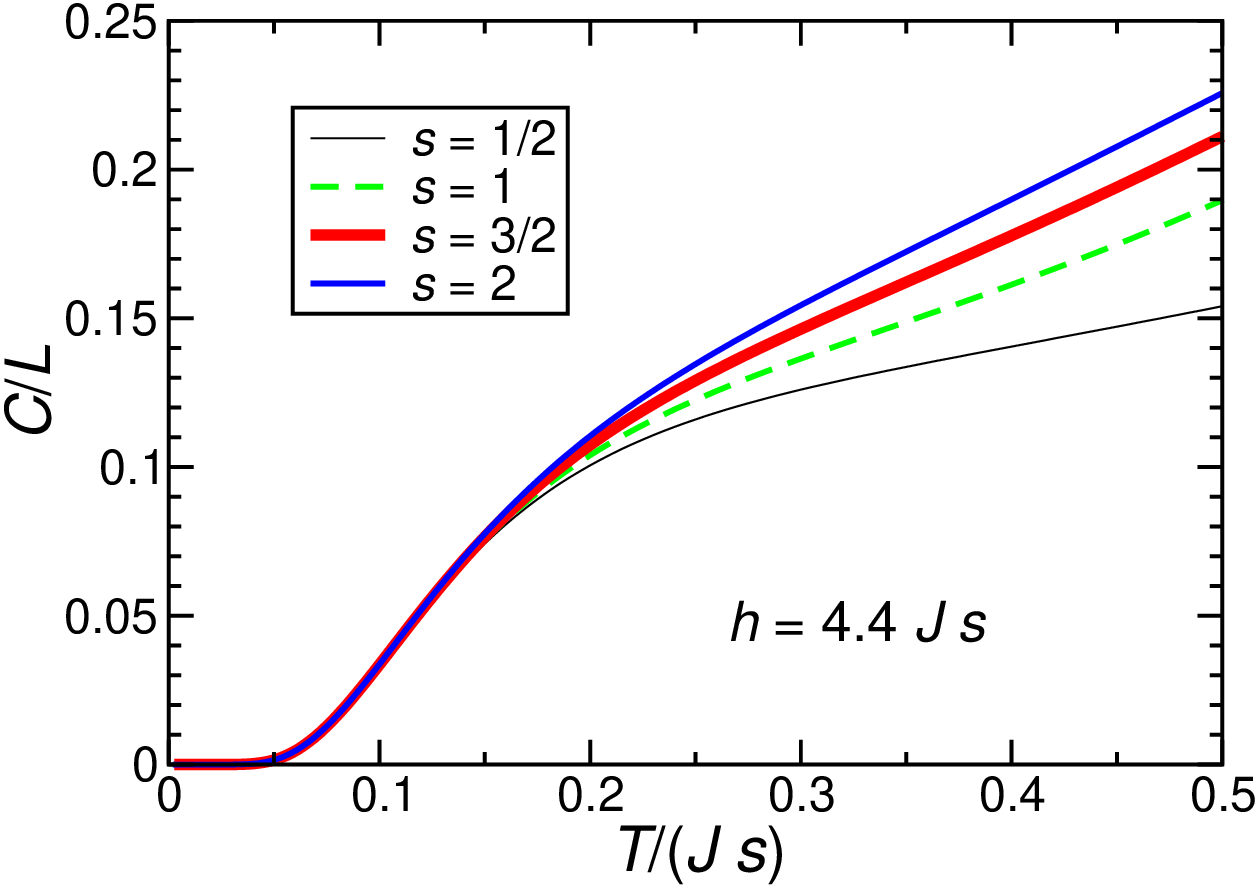}
\hfill
\includegraphics[width=0.48\columnwidth]{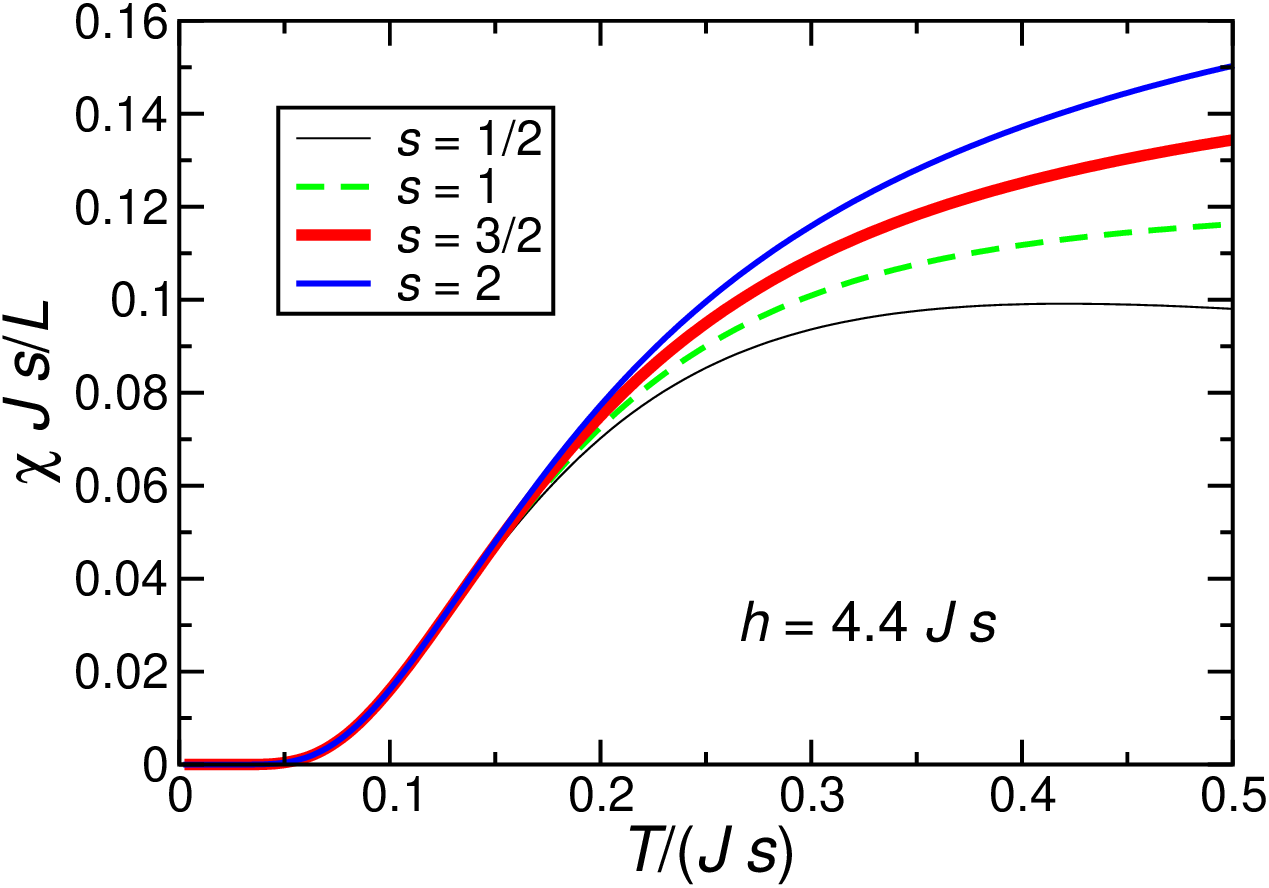} \\
\includegraphics[width=0.48\columnwidth]{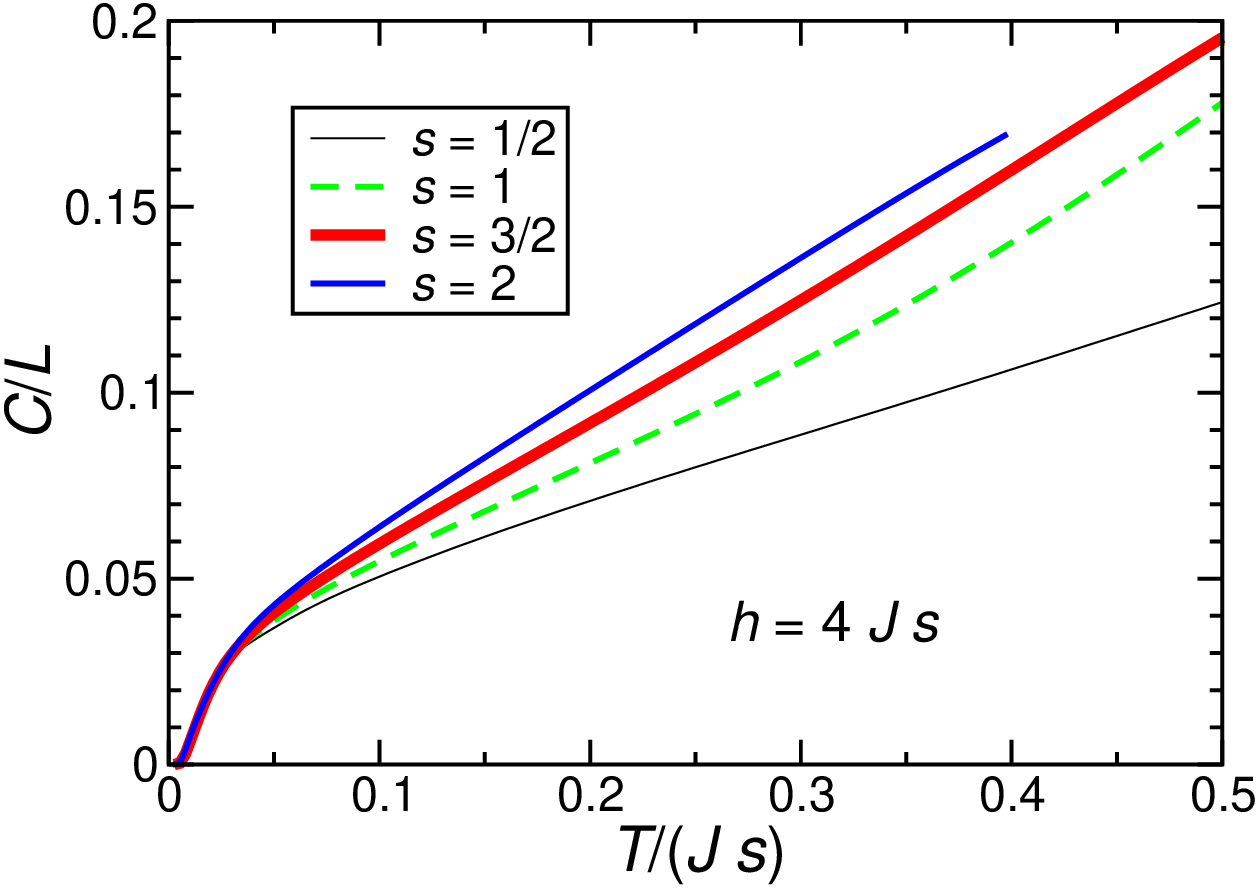}
\hfill
\includegraphics[width=0.48\columnwidth]{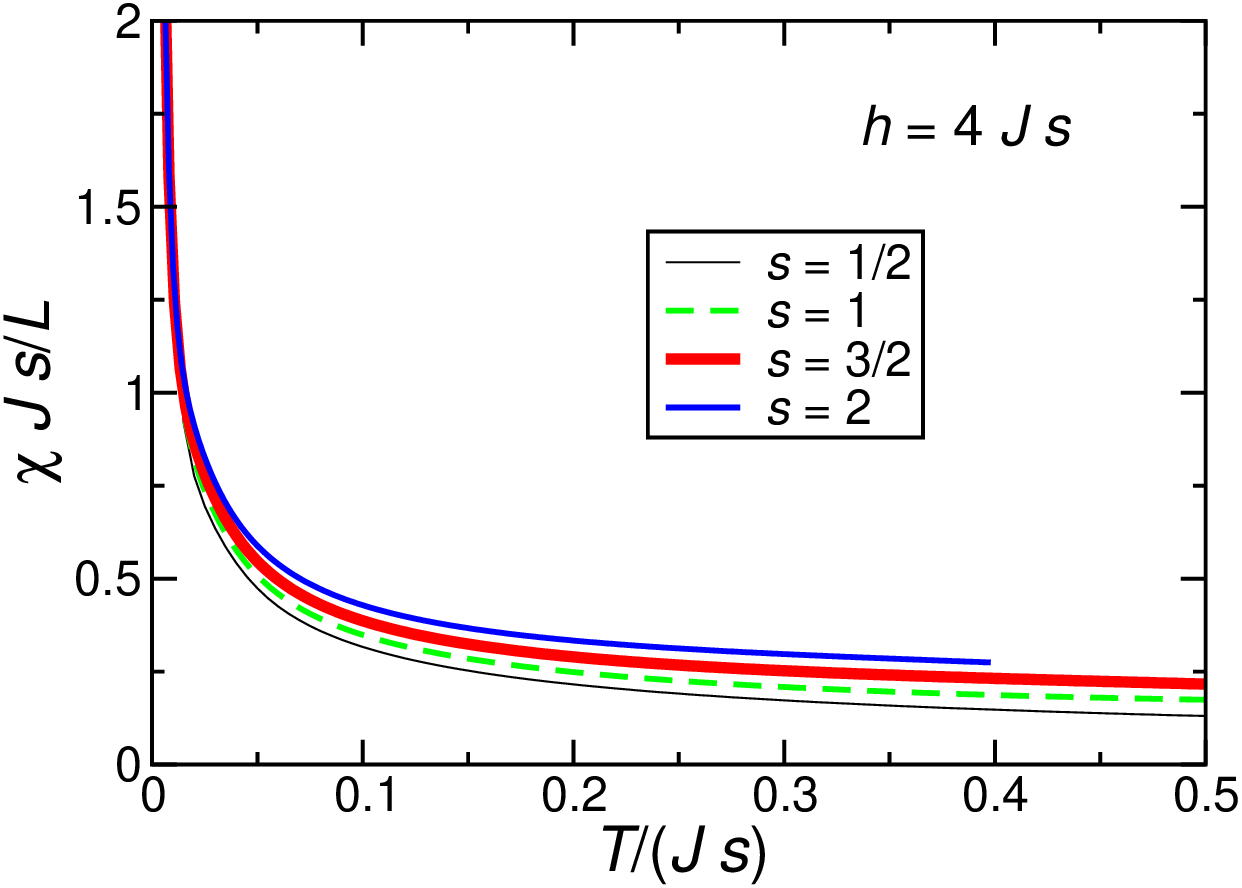} \\
\includegraphics[width=0.48\columnwidth]{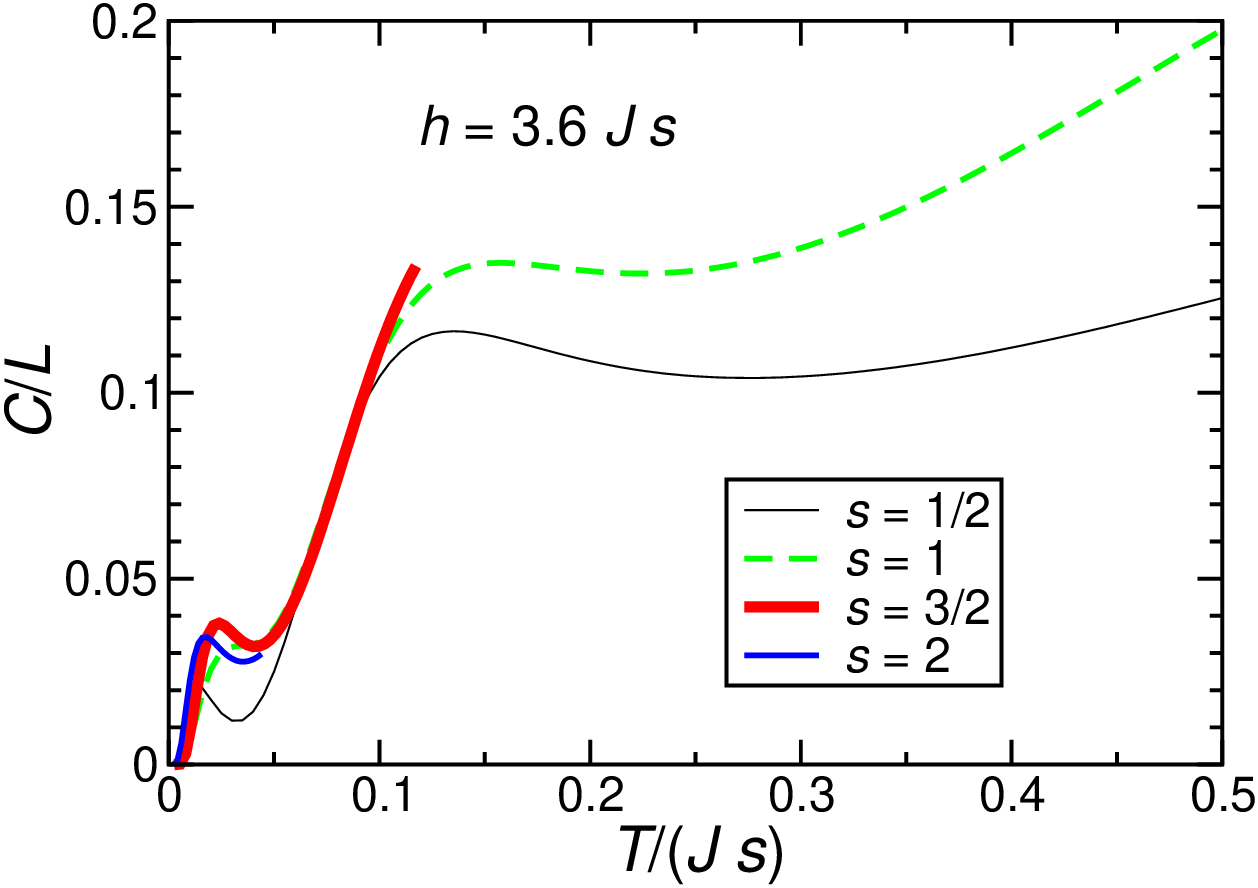}
\hfill
\includegraphics[width=0.48\columnwidth]{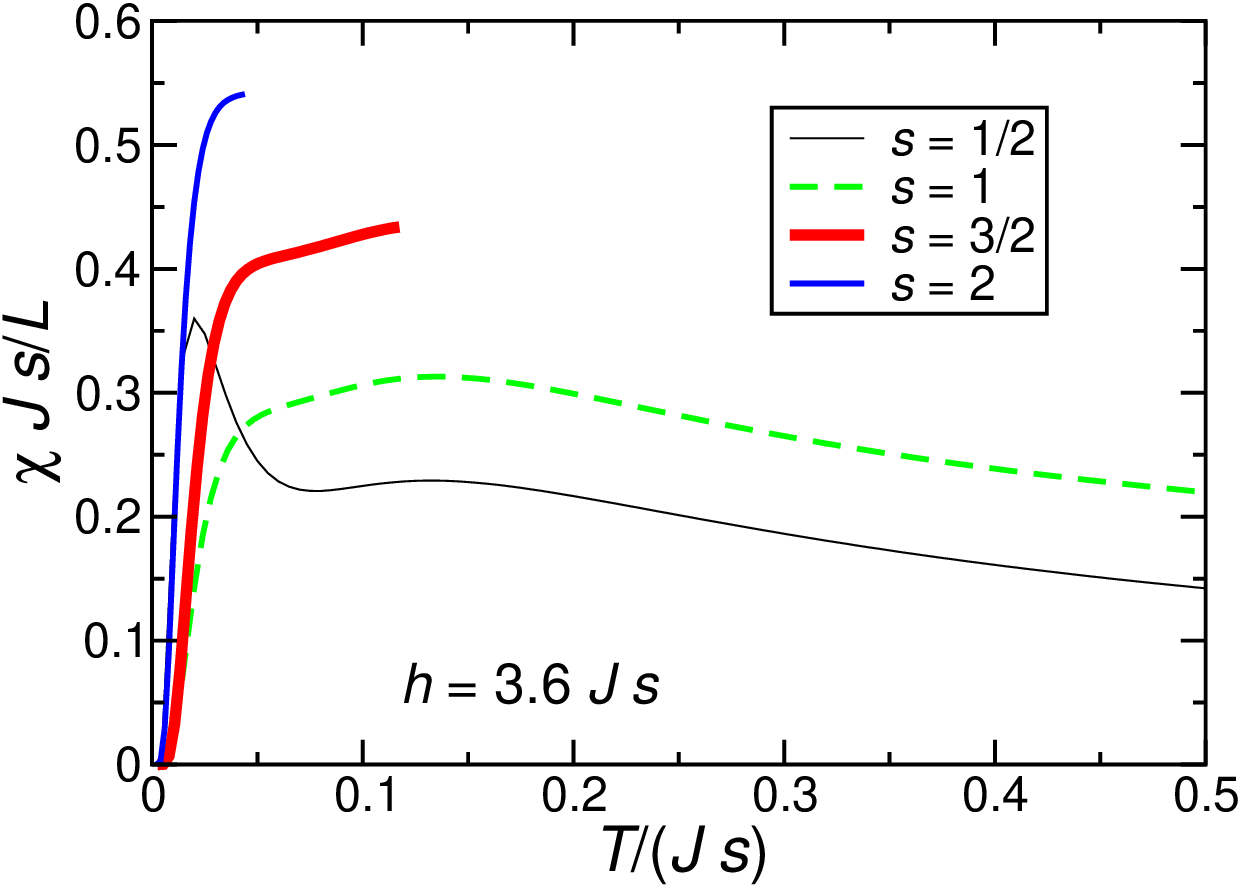}
\end{center}
\caption{Specific heat per site $C/L$ (left column) and magnetic
susceptibility per site $\chi/L$ (right column) as obtained
by ED for different $s$. Note that a truncation of the
energy spectrum has been used for $s > 1/2$. The values
of the magnetic field are chosen somewhat above the
saturation field (top row), exactly at the the saturation field
(middle row), and somewhat below the
saturation field (bottom row).}
\label{fig:scaleCchi}
\end{figure}

Further details can be read off from the temperature scans
at a fixed magnetic field shown in Fig.~\ref{fig:scaleCchi}. In
this figure, we have chosen to show the specific heat
$C$ as a representative for energy-related quantities\footnote{%
The specific heat $C=T\,(\partial\,S/\partial T)$ is equivalent
to $S$ since in the present case $S(0)/L \to 0$ for $L \to \infty$.
} and the magnetic susceptibility
$\chi$ as a representative for magnetic quantities.
The top row of  Fig.~\ref{fig:scaleCchi} corresponds to
a magnetic field $h = 1.1\,h_{\rm sat.}$. Here we
observe a nice collapse with $s$ for $T/(J\,s) \lesssim 0.15$.
The middle row of Fig.~\ref{fig:scaleCchi} corresponds to
a magnetic field exactly equal to the saturation field.
Here we observe a good scaling collapse for $T/(J\,s) \lesssim 0.04$.
Finally, the bottom row of Fig.~\ref{fig:scaleCchi} corresponds to
a magnetic field $h = 0.9\,h_{\rm sat.}$. In this case, one expects
the scaling region to be pushed to even lower temperatures.
However, for $h < h_{\rm sat.}$ the finite size $L=20$ leads to artifacts
in the low-temperature behavior. Therefore it is difficult to make
definite statements for this case.

To summarize this section, we have provided evidence of {\em linear}
scaling with $s$ for thermodynamic quantities of spin-$s$ Heisenberg
chains close to the saturation field. Note that this is very different from
the {\em quadratic} scaling, e.g.\ of $T$ with $s^2$ \cite{ZH09}
or with $s\,(s+1)$ \cite{KGWB98,ELS06},
which is needed to approach the classical limit.

\section{Conclusions and outlook}

\label{sec:Summary}

In this paper we have illustrated the numerical computation of thermodynamic
quantities and in particular the entropy for spin-$s$ Heisenberg chains. From
a technical point of view, we have described in section \ref{sec:ED} how
to perform a reliable computation of a large number of low-lying
states using the Lanczos method and in section \ref{sec:QMC} how to
compute the entropy directly by a QMC simulation.

In section \ref{sec:S1chain} we have then focused on the $s=1$
Heisenberg chain and shown in particular that one can cool with an
adiabatic magnetization process during which the Haldane gap $\Gap$ is closed.
Many previous investigations (e.g.,
\cite{Zhito,ZhiHo,WeHo,SST06,STS07,ZH09})
have focused on the saturation field. The reason for this
is that the saturation field is 
a field-induced quantum phase transition at a known value of
the magnetic field which also gives rise to technical simplifications.
However, the scenario of quantum phase transitions \cite{ZGRS03,GR05}
is universal and not restricted to the saturation field. Indeed,
Fig.~\ref{figS1entHgap} is consistent with the same universal
square-root behavior of the entropy
$S$ at $h=\Delta$ in the spin-1 Heisenberg chain as observed previously for
$s=1/2$ chains exactly at the saturation field \cite{BoNa72,ZhiHo}.
This may also be important from an experimental point of view since
a possible spin gap may be accessible by laboratory magnetic fields even
if the saturation field is out of reach. In fact,
cooling by adiabatic magnetization when closing a spin gap has presumably
been indirectly observed in pulse-field magnetization experiments on
SrCu$_2$(BO$_3$)$_2$ \cite{SCBO08}.

In section \ref{sec:Sscal} we then moved to the saturation field
and investigated scaling with the spin quantum number $s$.
In contradistinction to the classical scaling regime where temperature
should scale quadratically with $s$ \cite{ZH09,KGWB98,ELS06}, there is
a quantum scaling regime around the saturation field where one observes
a collapse using the linearly scaled parameters $h/s$ and $T/s$.
Also in higher dimensions the one-magnon dispersion typically scales
with $s$. By the same arguments as in section~\ref{sec:Sscal}, we therefore
expect linear scaling with $s$ at any continuous transition to saturation.
This expectation could be tested numerically with the methods of
the present paper.

\section*{Acknowledgments}

We are grateful to M.E.\ Zhitomirsky for useful discussions.
A.H.\ acknowledges support by the Deutsche Forschungsgemeinschaft through a
Heisenberg fellowship (Project HO~2325/4-1).
S.W.\ acknowledges HLRS Stuttgart and NIC J\"ulich for allocation of computing time.
Some of our numerical simulations were based on the ALPS libraries \cite{ALPS1,ALPS2}.

%
%
  \label{last@page}
  \end{document}